\documentclass[aps,pre,preprint,superscriptaddress]{revtex4}
\usepackage{graphics,epsfig,wasysym}

\begin{document}

\centerline{\large\bf The existence of a bending rigidity for a hard sphere liquid}
\centerline{\large\bf near a curved hard wall: Helfrich or Hadwiger?}
\vskip 20pt
\centerline{Edgar M. Blokhuis}
\vskip 5pt
\centerline{\it Colloid and Interface Science, Leiden Institute of Chemistry,}
\centerline{\it Gorlaeus Laboratories, P.O. Box 9502, 2300 RA Leiden, The Netherlands.}
\vskip 15pt
\centerline{\bf Abstract}
\vskip 5pt
\noindent
In the context of Rosenfeld's Fundamental Measure Theory, we show that
the bending rigidity is not equal to zero for a hard-sphere fluid in
contact with a curved hard wall. The implication is that the Hadwiger Theorem
does not hold in this case and the surface free energy is given by
the Helfrich expansion instead. The value obtained for the bending rigidity
is (1) an order of magnitude smaller than the bending constant associated
with Gaussian curvature, (2) changes sign as a function of the fluid volume
fraction, (3) is independent of the choice for the location of the hard wall.
\vskip 5pt
\noindent
\centerline{\rule{300pt}{1pt}}

\section{Introduction}

\noindent
The Helfrich free energy \cite{Helfrich} has proven to be an invaluable starting
point in the description of the surface properties of complex surfaces such as
membranes or surfactant systems \cite{surfactants, Safran}. It is the most
general form for the surface (or excess) free energy of an isotropic surface
expanded to second order in the surface's curvature:
\begin{equation}
\label{eq:Helfrich}
\Omega_{\rm H} = \int \!\! dA \; [ \, \sigma - \delta \sigma \, J
+ \frac{k}{2} \, J^2 + \bar{k} \, K \, + \ldots ] \,,
\end{equation}
where $J \!=\! 1/R_1 + 1/R_2$ is the total curvature, $K \!=\! 1/(R_1R_2)$
is the Gaussian curvature and $R_1$, $R_2$ are the principal radii of curvature
at a certain point on the surface. The expansion defines four curvature coefficients:
$\sigma$, the surface tension of the planar interface, $\delta$, the Tolman
length \cite{Tolman}, $k$, the bending rigidity, and $\bar{k}$, the rigidity
constant associated with Gaussian curvature. The original expression proposed
by Helfrich \cite{Helfrich} features the radius of spontaneous curvature $R_0$ as the
linear curvature term ($\delta \sigma \rightarrow 2 k / R_0$ \cite{Blokhuis92, Blokhuis06}),
but in honour of Tolman, who was the first to consider curvature corrections
to the surface tension \cite{Tolman}, we use the notation in Eq.(\ref{eq:Helfrich}).

Recently, an alternative description to replace the Helfrich free energy in certain
situations was put forward by K\"{o}nig {\em et al.} \cite{Konig04, Konig05} based
on the implications of the {\em Hadwiger Theorem} \cite{Hadwiger, Mecke}. The Hadwiger
Theorem states that any functional of a system that is translationally
invariant, additive and continuous, can be written as a linear combination
of the four Minkowski functionals: volume, surface area, and the integrated
total and Gaussian curvatures \cite{Mecke}. The implication is that, as an
alternative to Eq.(\ref{eq:Helfrich}), the surface free energy can be written as:
\begin{equation}
\label{eq:Hadwiger}
\Omega_{\rm Hadwiger} = \int \!\! dA \; [ \, \sigma - \delta \sigma \, J + \bar{k} \, K ] \,.
\end{equation}
Comparing the two expressions for the free energy in Eqs.(\ref{eq:Helfrich})
and (\ref{eq:Hadwiger}), we are led to the following two implications of
the Hadwiger Theorem:\\
{\bf 1.} The bending rigidity constant is zero,\\
{\bf 2.} Higher order curvature terms, represented by the dots in Eq.(\ref{eq:Helfrich}), are absent.\\

The question now is for which systems are the conditions of the Hadwiger
Theorem fulfilled so that the bending rigidity and higher order curvature
terms are all strictly zero? It was suggested that for a hard sphere
fluid in contact with a hard, structureless wall, the Hadwiger Theorem
should hold and Eq.(\ref{eq:Hadwiger}) is a {\em complete} expression
for its surface free energy \cite{Konig04, Konig05}. The evidence for
this suggestion is based on a numerical analysis \cite{Konig04, Konig05,
Bryk03} of the free energy in spherical and cylindrical geometry
using Rosenfeld's Fundamental Measure Theory (FMT) \cite{Rosenfeld, Roth},
showing that the bending rigidity is zero within numerical accuracy.
To understand the basis for this result in more detail, we consider surfaces
for which the curvatures $J$ and $K$ are constant. The Helfrich free energy per
unit area is then given by:
\begin{equation}
\label{eq:sigma(J,K)}
\Omega_{\rm H}/A \equiv \sigma(J,K) = \sigma - \delta \sigma \, J + \frac{k}{2} \, J^2 + \bar{k} \, K + \ldots
\end{equation}
For a spherically or cylindrically shaped surface with radius $R$,
this expansion then takes the form:
\begin{eqnarray}
\label{eq:sigma_s(R)}
\sigma_s(R) &=& \sigma - \frac{2 \delta \sigma}{R} 
+ \frac{(2 k + \bar{k})}{R^2} + \ldots \hspace*{27pt} {\rm (sphere)} \\
\label{eq:sigma_c(R)}
\sigma_c(R) &=& \sigma - \frac{\delta \sigma}{R} 
+ \frac{k}{2 R^2} + \ldots \hspace*{56pt} {\rm (cylinder)}
\end{eqnarray}
Note that only the combination $2 k + \bar{k}$ appears in the expression
for the surface tension of the spherical interface, so that the conclusion
whether $k$ is identically zero or not can be made only from an analysis
of the {\em cylindrical} system. Next to the curvature dependent surface tension,
one may also investigate the curvature dependence of the {\em wall density}
$\rho^{\rm W}$. According to the {\em wall theorem}, the wall density of a
fluid in contact with an infinitely hard, {\em planar} wall is related to
the bulk pressure $p$ through an ideal gas law \cite{Lebowitz, Henderson83}:
\begin{equation}
\label{eq:wall_theorem}
k_{\rm B} T \, \rho^{\rm W} = p \,.
\end{equation}
The wall theorem can be generalized to a spherically shaped hard wall
\cite{Henderson83, Henderson86, Blokhuis07} or to a more generally shaped hard wall
with (constant) curvatures $J$ and $K$ \cite{Blokhuis_unpublished}:
\begin{equation}
\label{eq:rho_w(J,K)}
k_{\rm B} T \, \rho^{\rm W}(J,K) = p + \sigma \, J - 2 \delta \sigma \, K
- \frac{k}{2} \, J^3 + 2 k \, J \, K + \ldots
\end{equation}
Note that a term proportional to $J^2$ is absent in the expression above \cite{Note}.
For a spherically or cylindrically shaped surface, this expansion takes the form:
\begin{eqnarray}
\label{eq:rho_s_w(R)}
k_{\rm B} T \, \rho_s^{\rm W}(R) &=& p + \frac{2 \sigma}{R} - \frac{2 \delta \sigma}{R^2}
+ \dots \hspace*{40pt} {\rm (sphere)} \\
\label{eq:rho_c_w(R)}
k_{\rm B} T \, \rho_c^{\rm W}(R) &=& p + \frac{\sigma}{R} - \frac{k}{2 R^3}
+ \dots \hspace*{43pt} {\rm (cylinder)}
\end{eqnarray}
where the dots represent terms of ${\cal O}(1/R^4)$ which indicates that the
term proportional to $1/R^3$ is absent in the expansion of the spherical
interface. The corresponding term in the expansion of the {\em cylindrical}
interface is related to the bending rigidity thus supplying a {\em second route}
to the determination of its value. Note that these expressions are valid only
when the radius $R$ is defined via the wall density
$\rho^{\rm W} \!\equiv\! \rho(r \!=\! R^+)$. 

In this article, we revisit the analysis by K\"{o}nig {\em et al.}
\cite{Konig04, Konig05} for a hard sphere fluid in contact with a hard wall.
Using the exact same theoretical model as in refs. \cite{Konig04, Konig05, Bryk03},
i.e. FMT \cite{Rosenfeld, Roth}, we show in Section \ref{sec-FMT} that a detailed numerical
analysis yields a bending rigidity that is {\bf not equal to zero}, but an order
of magnitude smaller than the rigidity constant associated with Gaussian curvature.
Consistent values for $k$ are obtained from the analysis of the radius dependence
of the surface tension, Eq.(\ref{eq:sigma_c(R)}), as well as from the
analysis of the radius dependence of the wall density, Eq.(\ref{eq:rho_c_w(R)}).

As a further consistency test, we perform a systematic expansion of the FMT
free energy to second order in the curvature for the spherical and cylindrical
interface in Section \ref{sec-expansion}. This expansion is analogous to a
similar expansion for the liquid-vapour interface \cite{Blokhuis93}.
It is shown that the resulting expressions for $\sigma$, $\delta$,
and $\bar{k}$ are all in terms of the fluid density profile of the {\em planar}
interface, $\rho_0(z)$, whereas the expression for the bending rigidity $k$,
features the leading order curvature correction to the density profile,
$\rho_1(z)$. The values obtained for $\sigma$, $\delta \sigma$, and the
combination $2 k + \bar{k}$ using these expressions are all consistent with
the results of K\"{o}nig {\em et al.} \cite{Konig04, Konig05} and those by
Bryk {\em et al.} \cite{Bryk03}. 
The value obtained for the {\em bending rigidity} is not zero and consistent with the
two values obtained from the radius dependent surface tension and wall density.
Furthermore, it is in {\em qualitative} agreement with recent MD simulations
by Laird {\em et al.} \cite{Laird12} who determined the curvature dependent
surface tension of a fluid near a hard wall by Gibbs-Cahn integration \cite{Laird12, Laird10}.

\section{Fundamental measure theory}
\label{sec-FMT}

\noindent
In this section, we discuss Rosenfeld's Fundamental Measure Theory \cite{Rosenfeld}
as it is applied specifically to a one-component fluid consisting of spherical
particles with a diameter $d$. The free energy is then the following functional
of the fluid density $\rho (\vec{r})$ \cite{Rosenfeld, Roth}:
\begin{equation}
\label{eq:Omega_FMT}
\frac{\Omega[\rho]}{k_{\rm B} T} = \int \!\! d\vec{r} \; 
\left[ \rho \, \ln(\rho) - \rho - \frac{\mu}{k_{\rm B} T} \, \rho
+ \frac{V_{\rm ext}(\vec{r})}{k_{\rm B} T} \, \rho + \phi \right] \,,
\end{equation} 
where $\mu$ is the chemical potential and where the external field
$V_{\rm ext}(\vec{r})$ is used to express the presence of the 
hard wall. For spherically shaped fluid particles the free energy
density $\phi \!=\! \phi(n_2, n_3, \vec{n}_V)$ is explicitly given by 
\begin{equation}
\phi = \frac{1}{\pi d^2} \left[ - n_2 \, \ln(1-n_3)
+ \frac{d^2 (n_2^2 - | \vec{n}_{\rm V} |^2)}{2 (1-n_3)}
+ \frac{d^2 (n_2^3 - 3 n_2 \, | \vec{n}_{\rm V} |^2)}{24 (1-n_3)^2} \right] \,.
\end{equation} 
The three densities $n_{\alpha}(\vec{r})$ ($\alpha \!=\! 2, 3, V$) are different
convolutions of the fluid density
\begin{equation}
\label{eq:n_FMT}
n_{\alpha}(\vec{r}_1) = \int \!\! d\vec{r}_2 \; \rho (\vec{r}_2) \, w_{\alpha}(\vec{r}_1-\vec{r}_2) \,,
\end{equation} 
where the weight functions $w_{\alpha}(\vec{r})$ are explicitly given by \cite{Roth}
\begin{equation}
w_2(\vec{r}) = \delta(\frac{d}{2}-r) \,, \hspace*{30pt}
w_3(\vec{r}) = \Theta(\frac{d}{2}-r) \hspace*{20pt} {\rm and} \hspace*{20pt}
\vec{w}_{\rm V}(\vec{r}) = \frac{\vec{r}}{r} \, \delta(\frac{d}{2}-r) \,.
\end{equation} 
The Euler-Lagrange equation that minimizes the free energy in
Eq.(\ref{eq:Omega_FMT}) is given by
\begin{equation}
\label{eq:EL_FMT}
\frac{\mu}{k_{\rm B} T} = \ln(\rho) + \frac{V_{\rm ext}(\vec{r})}{k_{\rm B} T} + \sum\limits_{\alpha} \int \!\! d\vec{r}_2 \;
\frac{\partial \phi}{\partial n_{\alpha}(\vec{r}_2)} \, w_{\alpha}(\vec{r}_2-\vec{r}_1) \,.
\end{equation} 
Note that the Euler-Lagrange equation features $w_{\alpha}(\vec{r}_2-\vec{r}_1)$
and not $w_{\alpha}(\vec{r}_1-\vec{r}_2)$ as in Eq.(\ref{eq:n_FMT}) \cite{Roth}.

For a {\em uniform} system, we have that $n_2 \!=\! 6 \eta / d$, $n_3 \!=\! \eta$
and $\vec{n}_{\rm V} \!=\! 0$, with the volume fraction defined as $\eta \!\equiv\! (\pi/6)
\, \rho \, d^3$. The Euler-Lagrange equation in Eq.(\ref{eq:EL_FMT}) then becomes:
\begin{equation}
\label{eq:mu}
\frac{\mu}{k_{\rm B} T} = \ln(\rho) - \ln(1-\eta) + \frac{\eta \, (14 -13 \eta + 5 \eta^2)}{2(1-\eta)^3} \,.
\end{equation} 
Using the expression for the chemical potential above the bulk pressure is obtained
from $\Omega \!=\! - p V$ giving the Percus-Yevick equation of state:
\begin{equation}
\label{eq:EOS}
\frac{d^3 \, p}{k_{\rm B} T} = \frac{6 \eta \, (1 + \eta + \eta^2)}{\pi(1-\eta)^3} \,.
\end{equation} 
We mention that a refinement of FMT was recently proposed \cite{FMT_CS} to yield
the more accurate Carnahan Starling equation of state \cite{CS} instead of Eq.(\ref{eq:EOS}).
It is expected that results do not depend sensitively on this refinement.

Next, we consider the implementation of FMT in three different geometries: the planar,
spherical, and cylindrical interface.
\vskip 10pt
\noindent
{\bf Planar interface}
\vskip 5pt
\noindent
In planar geometry, we can simplify the expressions for $n_{\alpha}(\vec{r}) \!=\! n_{\alpha}(z)$,
where $z$ is the coordinate normal to the interface, as:
\begin{equation}
\label{eq:n_planar}
n_{\alpha}(z_1) = \int \!\! dz_2 \; \rho(z_2) \, w^0_{\alpha}(z_1-z_2) \,,
\end{equation} 
where the weight functions $w^0_{\alpha}(z)$ are explicitly given by
\begin{eqnarray}
\label{eq:w^0}
w^0_2(z) &=& \pi d \, \Theta(\frac{d}{2}-|z|)  \,, \\
w^0_3(z) &=& \pi (\frac{d^2}{4} - z^2) \, \Theta(\frac{d}{2}-|z|) \,, \nonumber \\
w^0_{\rm V}(z) &=& 2 \pi z \, \Theta(\frac{d}{2}-|z|) \,, \nonumber
\end{eqnarray} 
and where $\vec{n}_{\rm V}(\vec{r}) \!=\! n_{\rm V}(z) \hat{z}$. The Euler-Lagrange
equation in Eq.(\ref{eq:EL_FMT}) simplifies in planar geometry to
\begin{equation}
\label{eq:EL_planar}
\frac{\mu}{k_{\rm B} T} = \ln(\rho) + \frac{V_{\rm ext}(z_1)}{k_{\rm B} T} + \sum\limits_{\alpha} \int \!\! dz_2 \;
\frac{\partial \phi}{\partial n_{\alpha}(z_2)} \, w^0_{\alpha}(z_2-z_1) \,.
\end{equation} 
The external field mimics the presence of a hard wall for $z \!<\! 0$,
i.e. $V_{\rm ext}(z) \!=\! \infty$ when $z<0$ and zero otherwise, so that
the density $\rho(z) \!=\! 0$ for $z \!<\! 0$. The surface tension is
the surface free energy per unit area ($\sigma \!=\! (\Omega + p \, V)/A$ \cite{RW}):
\begin{equation}
\label{eq:sigma_FMT}
\frac{\sigma}{k_{\rm B} T} = \int\limits_{-d/2}^{\infty} \!\!\!\!\! dz
\left[ \rho \, \ln(\rho) - \rho - \frac{\mu}{k_{\rm B} T} \, \rho + \phi
+ \frac{p}{k_{\rm B} T} \, \Theta(z) \right] \,,
\end{equation} 
where the lower integration reflects the fact that $\phi$ and the
convoluted densities $n_{\alpha}(z)$ are zero only when $z \!<\! -d/2$.
\vskip 10pt
\noindent
{\bf Spherical interface}
\vskip 5pt
\noindent
In spherical geometry, the densities $n_{\alpha}(\vec{r}) \!=\! n_{\alpha}(r)$,
with $r$ the radial distance, are:
\begin{eqnarray}
n_{2}(r_1) &=& \int \!\! dr_2 \left(\frac{r_2}{r_1}\right) \rho(r_2) \, w^s_{2}(r_1-r_2) \,, \\
n_{3}(r_1) &=& \int \!\! dr_2 \left(\frac{r_2}{r_1}\right) \rho(r_2) \, w^s_{3}(r_1-r_2) \,, \nonumber \\
n_{\rm V}(r_1) &=& \int \!\! dr_2 \left(\frac{r_2}{r_1}\right) \rho(r_2) \, w^s_{\rm V}(r_1-r_2)
+ \frac{1}{r_1} \int \!\! dr_2 \left(\frac{r_2}{r_1}\right) \rho(r_2) \, w^s_{3}(r_1-r_2) \,, \nonumber 
\end{eqnarray} 
where the weight functions are equal to those in planar geometry (Eq.(\ref{eq:w^0})):
\begin{eqnarray}
w^s_2(r_1-r_2)       &=& \pi d \, \Theta(\frac{d}{2}-|r_1 - r_2|)  \,, \\
w^s_{3}(r_1-r_2)     &=& \pi (\frac{d^2}{4} - (r_1-r_2)^2) \, \Theta(\frac{d}{2}-|r_1-r_2|) \,, \nonumber \\
w^s_{\rm V}(r_1-r_2) &=& 2 \pi (r_1 - r_2) \, \Theta(\frac{d}{2}-|r_1 - r_2|) \,, \nonumber
\end{eqnarray}
and where $\vec{n}_{\rm V}(\vec{r}) \!=\! n_{\rm V}(r) \hat{r}$. The Euler-Lagrange
equation in Eq.(\ref{eq:EL_FMT}) now reduces to
\begin{eqnarray}
\label{eq:EL_sphere}
\frac{\mu}{k_{\rm B} T} &=& \ln(\rho) + \frac{V_{\rm ext}(r_1)}{k_{\rm B} T} + \sum\limits_{\alpha} \int \!\! dr_2
\left(\frac{r_2}{r_1}\right) \frac{\partial \phi}{\partial n_{\alpha}(r_2)} \, w^s_{\alpha}(r_2-r_1) \\
&& + \frac{1}{r_1} \int \!\! dr_2 \; \frac{\partial \phi}{\partial n_{\rm V}(r_2)} \, w^s_3(r_2-r_1) \,. \nonumber 
\end{eqnarray} 
Again, the external field mimics the presence of a hard wall, i.e.
$V_{\rm ext}(r) \!=\! \infty$ when $r<R$, which serves to define the location
of the radius $R$ of the spherically shaped hard wall. The surface
tension now becomes:
\begin{equation}
\label{eq:sigma_sphere}
\frac{\sigma_s(R)}{k_{\rm B} T} = \int\limits_{R-d/2}^{\infty} \!\!\!\!\!\! dr \left(\frac{r}{R} \right)^{\!2}
\left[ \rho \, \ln(\rho) - \rho - \frac{\mu}{k_{\rm B} T} \, \rho + \phi
+ \frac{p}{k_{\rm B} T} \, \Theta(r-R) \right] \,.
\end{equation} 
\vskip 10pt
\noindent
{\bf Cylindrical interface}
\vskip 5pt
\noindent
In cylindrical geometry, the densities $n_{\alpha}(\vec{r}) \!=\! n_{\alpha}(r)$,
with $r$ the radial distance to the cylinder axis, reduces to:
\begin{eqnarray}
n_{2}(r_1) &=& \int \!\! dr_2 \left(\frac{r_2}{r_1}\right)^{\!\frac{1}{2}} \rho(r_2) \, w^c_{2}(r_1,r_2) \,, \\
n_{3}(r_1) &=& \int \!\! dr_2 \left(\frac{r_2}{r_1}\right)^{\!\frac{1}{2}} \rho(r_2) \, w^c_{3}(r_1,r_2) \,, \nonumber \\
n_{\rm V}(r_1) &=& \int \!\! dr_2 \left(\frac{r_2}{r_1}\right)^{\!\frac{1}{2}} \rho(r_2) \, w^c_{\rm V}(r_1,r_2)
+ \frac{1}{2 r_1} \int \!\! dr_2 \left(\frac{r_2}{r_1}\right)^{\!\frac{1}{2}} \rho(r_2) \, w^c_{3'}(r_1,r_2) \,, \nonumber
\end{eqnarray} 
where the weight functions are given by:
\begin{eqnarray}
\label{eq:w^c}
w^c_2(r_1,r_2)       &=& 2 d \, K(\beta) \, \Theta(\frac{d}{2}-|r_2-r_1|)  \,, \\
w^c_{3}(r_1,r_2)     &=& 16 \, r_1 r_2 \, [ E(\beta) + (\beta^2-1) K(\beta)] \, \Theta(\frac{d}{2}-|r_2-r_1|) \,, \nonumber \\
w^c_{3'}(r_1,r_2)    &=& 16 \, r_1 r_2 \, [ K(\beta) - E(\beta)] \, \Theta(\frac{d}{2}-|r_2-r_1|) \,, \nonumber \\
w^c_{\rm V}(r_1,r_2) &=& 4 (r_1 - r_2) K(\beta) \, \Theta(\frac{d}{2}-|r_2-r_1|) \,, \nonumber
\end{eqnarray}
where $\vec{n}_{\rm V}(\vec{r}) \!=\! n_{\rm V}(r) \hat{r}$ and where $K(\beta)$
and $E(\beta)$ are complete elliptic integrals of the first and second kind,
respectively \cite{Elliptic}. The argument of the elliptic functions is defined
as $\beta^2 \!\equiv\! [d^2/4 - (r_2 - r_1)^2] / (4 r_1 r_2)$. Note that the weight
functions in the cylindrical case are functions of the radial distances $r_1$ and
$r_2$, separately and not only the difference $r_1 - r_2$. The Euler-Lagrange equation
in Eq.(\ref{eq:EL_FMT}) in cylindrical geometry reduces to
\begin{eqnarray}
\label{eq:EL_cylinder}
\frac{\mu}{k_{\rm B} T} &=& \ln(\rho) + \frac{V_{\rm ext}(r_1)}{k_{\rm B} T} + \sum\limits_{\alpha} \int \!\! dr_2
\left(\frac{r_2}{r_1}\right)^{\!\frac{1}{2}} \frac{\partial \phi}{\partial n_{\alpha}(r_2)} \, w^c_{\alpha}(r_2,r_1) \\
&& + \frac{1}{2 r_1} \int \!\! dr_2 \left(\frac{r_1}{r_2}\right)^{\!\frac{1}{2}}
\frac{\partial \phi}{\partial n_{\rm V}(r_2)} \, w^c_{3'}(r_2,r_1) \,. \nonumber
\end{eqnarray} 
Again, the external field mimics the presence of a hard wall for $r \!<\! R$.
The surface tension in cylindrical geometry is given by:
\begin{equation}
\label{eq:sigma_cylinder}
\frac{\sigma_c(R)}{k_{\rm B} T} = \int\limits_{R-d/2}^{\infty} \!\!\!\!\!\! dr
\left( \frac{r}{R} \right) \left[ \rho \, \ln(\rho) - \rho - \frac{\mu}{k_{\rm B} T} \, \rho
+ \phi + \frac{p}{k_{\rm B} T} \, \Theta(r-R) \right] \,.
\end{equation} 

The procedure to evaluate $\sigma_s(R)$ and $\sigma_c(R)$ is now as
follows. For a certain fixed value of the fluid volume fraction $\eta$,
the corresponding chemical potential and pressure are determined from 
Eqs.(\ref{eq:mu}) and (\ref{eq:EOS}). Next, a value for the radius $R$
is chosen and the Euler-Lagrange equation in Eq.(\ref{eq:EL_sphere}) or
(\ref{eq:EL_cylinder}) is solved numerically to obtain the density profile
$\rho(r)$ (for details, see the excellent review on FMT by Roth in \cite{Roth}).
The density profile thus obtained then directly provides the wall density
$\rho^{\rm W} \!\equiv\! \rho(r \!=\! R^+)$ and the radius dependent
surface tension by evaluating the integral in Eq.(\ref{eq:sigma_sphere})
or (\ref{eq:sigma_cylinder}). Finally, the curvature coefficients are
obtained from a fit of the surface tension and wall density plotted as
a function of $1/R$ and comparing with the expansion in
Eqs.(\ref{eq:sigma_s(R)}) and (\ref{eq:sigma_c(R)}) or
Eqs.(\ref{eq:rho_s_w(R)}) and (\ref{eq:rho_c_w(R)}). The fit is carried
out by varying the reciprocal radius from 0 to 0.1 in steps of 0.01
in units of $1/d$. The resulting 11 data points are then fitted
(least-square) to polynomials in $1/R$ of progressing order starting from
a quadratic polynomial to a polynomial of order 7. It is verified that
the coefficients in the fit level off with the variation used as an
indication of the numerical error.

For the spherical interface, the polynomial fit of $\sigma_s(R)$ provides
values for the coefficients $\sigma$, $\delta \sigma$, and the combination
$2 k + \bar{k}$. The results are listed for three fluid volume fractions
in Table \ref{tb:Table1}.
\begin{table}
\centering
\begin{tabular}{| c || c | c | c | c | c |}
\hline
& $\sigma$ & \multicolumn{2}{c |}{$\delta \sigma$} & \multicolumn{2}{c |}{$2k + \bar{k}$} \\
$\eta$ & Eq.(\ref{eq:sigma}) & $\sigma_s(R)$ & Eq.(\ref{eq:delta}) & $\sigma_s(R)$ & Eq.(\ref{eq:k_sph}) \\
\hline
\hline
\hspace*{3pt} 0.10 \hspace*{3pt} & \hspace*{3pt} -0.0220978  \hspace*{3pt} 
& \hspace*{3pt} -0.00130941  \hspace*{3pt} & \hspace*{3pt} -0.00130941 \hspace*{3pt}
& \hspace*{3pt}  0.000428974 \hspace*{3pt} & \hspace*{3pt}  0.000428971 \hspace*{3pt} \\

\hspace*{3pt} 0.20 \hspace*{3pt} & \hspace*{3pt} -0.1394516 \hspace*{3pt}
& \hspace*{3pt} -0.014812  \hspace*{3pt} & \hspace*{3pt} -0.014811 \hspace*{3pt}
& \hspace*{3pt} -0.0014205 \hspace*{3pt} & \hspace*{3pt} -0.0014207 \hspace*{3pt} \\

\hspace*{3pt} 0.30 \hspace*{3pt} & \hspace*{3pt} -0.512482 \hspace*{3pt}
& \hspace*{3pt} -0.07321   \hspace*{3pt} & \hspace*{3pt} -0.07318 \hspace*{3pt}
& \hspace*{3pt} -0.0161    \hspace*{3pt} & \hspace*{3pt} -0.0161 \hspace*{3pt} \\
\hline
\end{tabular}
\vskip 20pt
\noindent
\caption{Numerical values for the surface tension $\sigma$ (in units of $k_{\rm B} T / d^2$),
Tolman length $\delta \sigma$ (in units of $k_{\rm B} T / d$) and the combination $2k + \bar{k}$
(in units of $k_{\rm B} T$) for three values of the volume fraction $\eta$. These values are
determined from an analysis of the radius dependence of the surface tension of a spherical
interface and by a direct evaluation of the expression in Eqs.(\ref{eq:sigma})-(\ref{eq:k_sph}).}
\label{tb:Table1}
\end{table}
The values for $\sigma$ and $\delta \sigma$ obtained from the polynomial
fit of the {\em wall density} are, within error, equal to those listed in the Table.

For the cylindrical interface, the polynomial fit of $\sigma_c(R)$ again
provides values for the coefficients $\sigma$ and $\delta \sigma$ (which
are consistent with the results in Table \ref{tb:Table1}), but the
coefficient of the $1/R^2$-term now yields values for the {\em rigidity
constant} $k$. These values are {\bf not equal to zero} within numerical accuracy
and are listed separately in Table \ref{tb:Table2}. Also listed are the
values obtained from the polynomial fit of the {\em wall density}.
Already it is noted that these two approaches are consistent and lead
to the conclusion that the bending rigidity is not equal to zero for
this system. To further corroborate this result, we consider a third
approach in the next section.

\begin{table}
\centering
\begin{tabular}{| c || c | c | c |}
\hline
& \multicolumn{3}{c |}{bending rigidity $k$ determined via:} \\
$\eta$ & $\sigma_c(R)$ & $\rho_c^{\rm W}(R)$ & Eq.(\ref{eq:rigidity}) \\
\hline
\hline
\hspace*{3pt} 0.10 \hspace*{3pt} & \hspace*{3pt}  0.000415172 \hspace*{3pt} & \hspace*{3pt} 0.000415165 \hspace*{3pt} & \hspace*{3pt} 0.000415171 \hspace*{3pt} \\

\hspace*{3pt} 0.20 \hspace*{3pt} & \hspace*{3pt}  0.00074260 \hspace*{3pt} & \hspace*{3pt} 0.00074251 \hspace*{3pt} & \hspace*{3pt} 0.00074254 \hspace*{3pt} \\

\hspace*{3pt} 0.30 \hspace*{3pt} & \hspace*{3pt} -0.000685 \hspace*{3pt} & \hspace*{3pt} -0.000615 \hspace*{3pt} & \hspace*{3pt} -0.000619 \hspace*{3pt} \\
\hline
\end{tabular}
\vskip 20pt
\noindent
\caption{Numerical values for the bending rigidity $k$ (in units of $k_{\rm B} T$)
for three values of the volume fraction $\eta$. The bending rigidity is determined
in three different ways: by an analysis of the radius dependence of the
{\em surface tension} of a cylindrical interface, the radius dependence
of the fluid {\em wall density} of a cylindrical interface, and by a direct
evaluation of the expression in Eq.(\ref{eq:rigidity}).}
\label{tb:Table2}
\end{table}

\noindent
\section{Curvature expansion}
\label{sec-expansion}

\noindent
In this section, we expand the free energy of the spherical and
cylindrical surface systematically to second order in $1/R$.
The analysis is outlined explicitly for the spherical interface --
the analysis of the cylindrical interface is more or less analogous,
but we indicate where it differs from that of the sphere.
\vskip 10pt
\noindent
{\bf Spherical interface}
\vskip 5pt
\noindent
All quantities are expanded to second order in the curvature. In particular,
the expansion of the density $\rho_s(r)$ reads:
\begin{equation}
\label{eq:expansion_rho}
\rho_s(r) = \rho_0(z) + \frac{\rho_1(z)}{R} + \frac{\rho_2(z)}{R^2} + \ldots \,,
\end{equation} 
where $z \!\equiv\!r-R$. The coefficients in the curvature expansion of the
density are determined from the curvature expansion of the Euler-Lagrange
equation in Eq.(\ref{eq:EL_sphere}). The result is that the (planar) density
profile $\rho_0(z)$ is determined from Eq.(\ref{eq:EL_planar}) and $\rho_1(z)$
follows from solving:
\begin{eqnarray}
\label{eq:EL_1}
0 &=& \frac{\rho_1(z_1)}{\rho_0(z_1)} 
+ \sum\limits_{\alpha, \beta} \int \!\! dz_2 \; \frac{\partial^2 \phi_0}{\partial n^0_{\alpha}(z_2)
\partial n^0_{\beta}(z_2)} \, n^1_{\beta}(z_1) \, w^0_{\alpha}(z_2-z_1) \\
&& + \sum\limits_{\alpha} \int \!\! dz_2 \; \frac{\partial \phi_0}{\partial n_{\alpha}(z_2)} \, z_{12} \, w^0_{\alpha}(z_2-z_1)
+ \int \!\! dz_2 \; \frac{\partial \phi_0}{\partial n_{\rm V}(z_2)} \, w^0_{3}(z_2-z_1) \,, \nonumber
\end{eqnarray} 
where $\phi_0 \!=\! \phi(\{n^0_{\alpha}\})$ and where we have defined $z_{12} \!\equiv\! z_2 - z_1$.
As we show below, it turns out that for the evaluation of the curvature
coefficients it is sufficient to obtain the density profiles $\rho_0(z)$
and $\rho_1(z)$ only. Using the expanded density profile, we can then
determine the coefficients in the expansion of $n^s_{\alpha}(r)$:
\begin{equation}
\label{eq:expansion_n}
n^s_{\alpha}(r) = n^0_{\alpha}(z) + \frac{n^1_{\alpha}(z)}{R} + \frac{n^2_{\alpha}(z)}{R^2} + \ldots \,,
\end{equation} 
where $n^0_{\alpha}(z)$ is given by Eq.(\ref{eq:n_planar}) and
$n^1_{\alpha}(z)$ can be calculated from
\begin{eqnarray}
n^1_{2}(z_1) &=& \int \!\! dz_2 \; \left[ \rho_1(z_2) + z_{12} \, \rho_0(z_2) \right] \, w^0_{2}(z_1-z_2) \,, \\
n^1_{3}(z_1) &=& \int \!\! dz_2 \; \left[ \rho_1(z_2) + z_{12} \, \rho_0(z_2) \right] \, w^0_{3}(z_1-z_2) \,, \nonumber \\
n^1_{\rm V}(z_1) &=& \int \!\! dz_2 \; \left\{ \left[ \rho_1(z_2) + z_{12} \, \rho_0(z_2) \right] \, w^0_{3}(z_1-z_2)
+ \rho_0(z_2) \, w^0_{3}(z_1-z_2) \right\} \,. \nonumber 
\end{eqnarray} 
Again, the evaluation of $n^2_{\alpha}(z)$ tuns out not to be necessary.

The expansions for $\rho_s(r)$ and $n^s_{\alpha}(r)$ are inserted into
the expression for the surface tension in Eq.(\ref{eq:sigma_sphere}).
Making a systematic expansion to second order in $1/R$, using the Euler-Lagrange
equations in Eqs.(\ref{eq:EL_planar}) and (\ref{eq:EL_1}), one ultimately
obtains expressions for the curvature coefficients by comparing to the
curvature expansion in Eq.(\ref{eq:sigma_s(R)}). For the surface tension of
the planar interface the result in Eq.(\ref{eq:sigma_FMT}) is recovered:
\begin{equation}
\label{eq:sigma}
\frac{\sigma}{k_{\rm B} T} = \int\limits_{-d/2}^{\infty} \!\!\!\! dz
\left[ \rho_0 \, \ln(\rho_0) - \rho_0 - \frac{\mu}{k_{\rm B} T} \, \rho_0 + \phi_0
+ \frac{p}{k_{\rm B} T} \, \Theta(z) \right] \,.
\end{equation} 
For the Tolman length one obtains
\begin{eqnarray}
\label{eq:delta}
\frac{\delta \sigma}{k_{\rm B} T} &=& - \int\limits_{-d/2}^{\infty} \!\!\!\! dz \; 
z \, \left[ \rho_0 \, \ln(\rho_0) - \rho_0 - \frac{\mu}{k_{\rm B} T} \, \rho_0 + \phi_0
+ \frac{p}{k_{\rm B} T} \, \Theta(z) \right] \\
&& - \frac{1}{2} \sum\limits_{\alpha} \int\limits_{-d/2}^{\infty} \!\!\!\! dz_1 \int\limits_{0}^{\infty} \!\! dz_2 \;
\frac{\partial \phi_0}{\partial n^0_{\alpha}(z_1)}
\, \rho_0(z_2) \, z_{12} \, w^0_{\alpha}(z_1-z_2) \nonumber \\
&& - \frac{1}{2} \int\limits_{-d/2}^{\infty} \!\!\!\! dz_1 \int\limits_{0}^{\infty} \!\! dz_2 \;
\frac{\partial \phi_0}{\partial n^0_{\rm V}(z_1)} \, \rho_0(z_2) \, w^0_3(z_1-z_2) \,. \nonumber
\end{eqnarray} 
For the combination $2 k + \bar{k}$ one finds
\begin{eqnarray}
\label{eq:k_sph}
\frac{2k+\bar{k}}{k_{\rm B} T} &=& \int\limits_{-d/2}^{\infty} \!\!\!\! dz \; 
z^2 \, \left[ \rho_0 \, \ln(\rho_0) - \rho_0 - \frac{\mu}{k_{\rm B} T} \, \rho_0 + \phi_0
+ \frac{p}{k_{\rm B} T} \, \Theta(z) \right] \\
&& + \sum\limits_{\alpha} \int\limits_{-d/2}^{\infty} \!\!\!\! dz_1 \int\limits_{0}^{\infty} \!\! dz_2 \;
z_1 \, \frac{\partial \phi_0}{\partial n^0_{\alpha}(z_1)}
\, \rho_0(z_2) \, z_{12} \, w^0_{\alpha}(z_1-z_2) \nonumber \\
&& - \frac{1}{2} \sum\limits_{\alpha} \int\limits_{-d/2}^{\infty}
\!\!\!\! dz_1 \int\limits_{0}^{\infty} \!\! dz_2 \; \frac{\partial \phi_0}{\partial n^0_{\alpha}(z_1)}
\, \rho_1(z_2) \, z_{12} \, w^0_{\alpha}(z_1-z_2) \nonumber \\
&& + \frac{1}{2} \sum\limits_{\alpha, \beta} \int\limits_{-d/2}^{\infty}
\!\!\!\! dz_1 \int\limits_{0}^{\infty} \!\! dz_2 \; \frac{\partial^2 \phi_0}{\partial n^0_{\alpha}(z_1) \partial n^0_{\beta}(z_1)}
\, n^1_{\beta}(z_1) \, \rho_0(z_2) \, z_{12} \, w^0_{\alpha}(z_1-z_2) \nonumber \\
&& + \frac{1}{2} \sum\limits_{\beta} \int\limits_{-d/2}^{\infty}
\!\!\!\! dz_1 \int\limits_{0}^{\infty} \!\! dz_2 \; \frac{\partial^2 \phi_0}{\partial n^0_{\rm V}(z_1) \partial n^0_{\beta}(z_1)}
\, n^1_{\beta}(z_1) \, \rho_0(z_2) \, w^0_3(z_1-z_2) \nonumber \\
&& + \frac{1}{2} \int\limits_{-d/2}^{\infty}
\!\!\!\! dz_1 \int\limits_{0}^{\infty} \!\! dz_2 \; \frac{\partial \phi_0}{\partial n^0_{\rm V}(z_1)}
\, \rho_1(z_2) \, w^0_3(z_1-z_2) \nonumber \\
&& + \int\limits_{-d/2}^{\infty}
\!\!\!\! dz_1 \int\limits_{0}^{\infty} \!\! dz_2 \; \frac{\partial \phi_0}{\partial n^0_{\rm V}(z_1)}
\, \rho_0(z_2) \, z_2 \, w^0_3(z_1-z_2) \,. \nonumber
\end{eqnarray} 

By solving the density profile $\rho_0(z)$ from Eq.(\ref{eq:EL_planar})
and $\rho_1(z)$ from Eq.(\ref{eq:EL_1}), these coefficients can all be
evaluated directly without having to determine the full radius dependent
surface tension as a function of $1/R$. It is therefore no surprise that
this route to the evaluation of the curvature coefficients is much more
convenient. To compare our results to the results by Bryk {\em et al.}
listed in their Table I \cite{Bryk03}, we need to take care of the fact that
in their analysis the location of the radius is defined according to the location
of the ``actual surface'' which accounts for the fact that the molecule's center
of mass is half a diameter away from the surface when it interacts with
the hard wall, $R_{\rm actual} \!=\! R - d/2$. The curvature coefficients
are then shifted according to the following transformation:
\begin{eqnarray}
\label{eq:transformation}
\left( \sigma \right)_{\rm R-d/2} &=& \sigma + \frac{p \, d}{2} \,, \\
\left( \delta \sigma \right)_{\rm R-d/2} &=& \delta \sigma - \frac{p \, d^2}{8} - \frac{\sigma \, d }{2} \,, \nonumber \\
\left( 2k + \bar{k} \right)_{\rm R-d/2} &=& 2k + \bar{k} + \frac{p \, d^3}{24} + \frac{\sigma \, d^2}{4} - \delta \sigma \, d \,. \nonumber
\end{eqnarray} 
The form of these transformations are derived by shifting the location
of the $z \!=\! 0$ plane by a distance $d/2$ in the expressions in
Eqs.(\ref{eq:sigma})-(\ref{eq:k_sph}).

\begin{figure}
\centering
\includegraphics[angle=270,width=250pt]{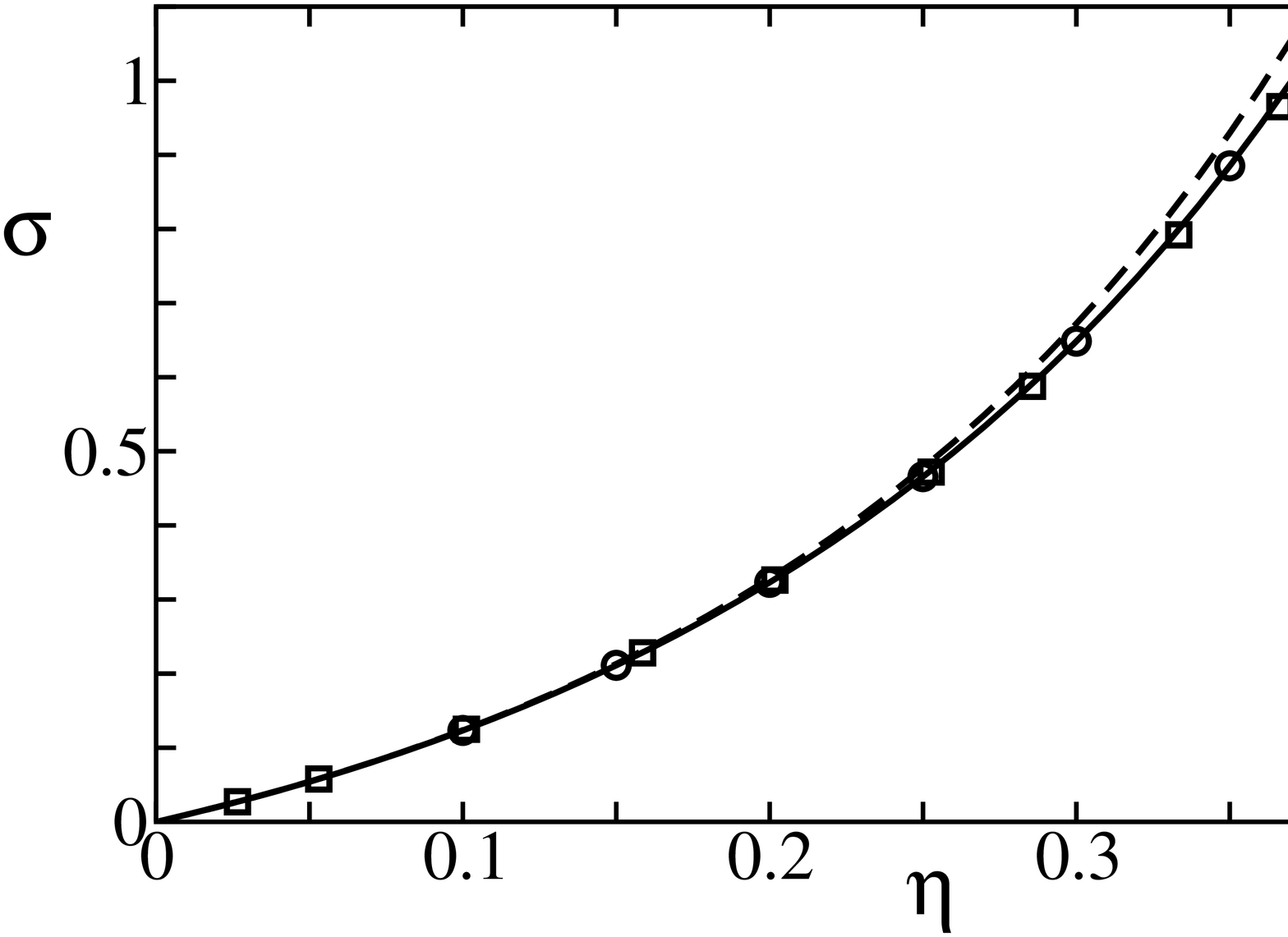}

\includegraphics[angle=270,width=250pt]{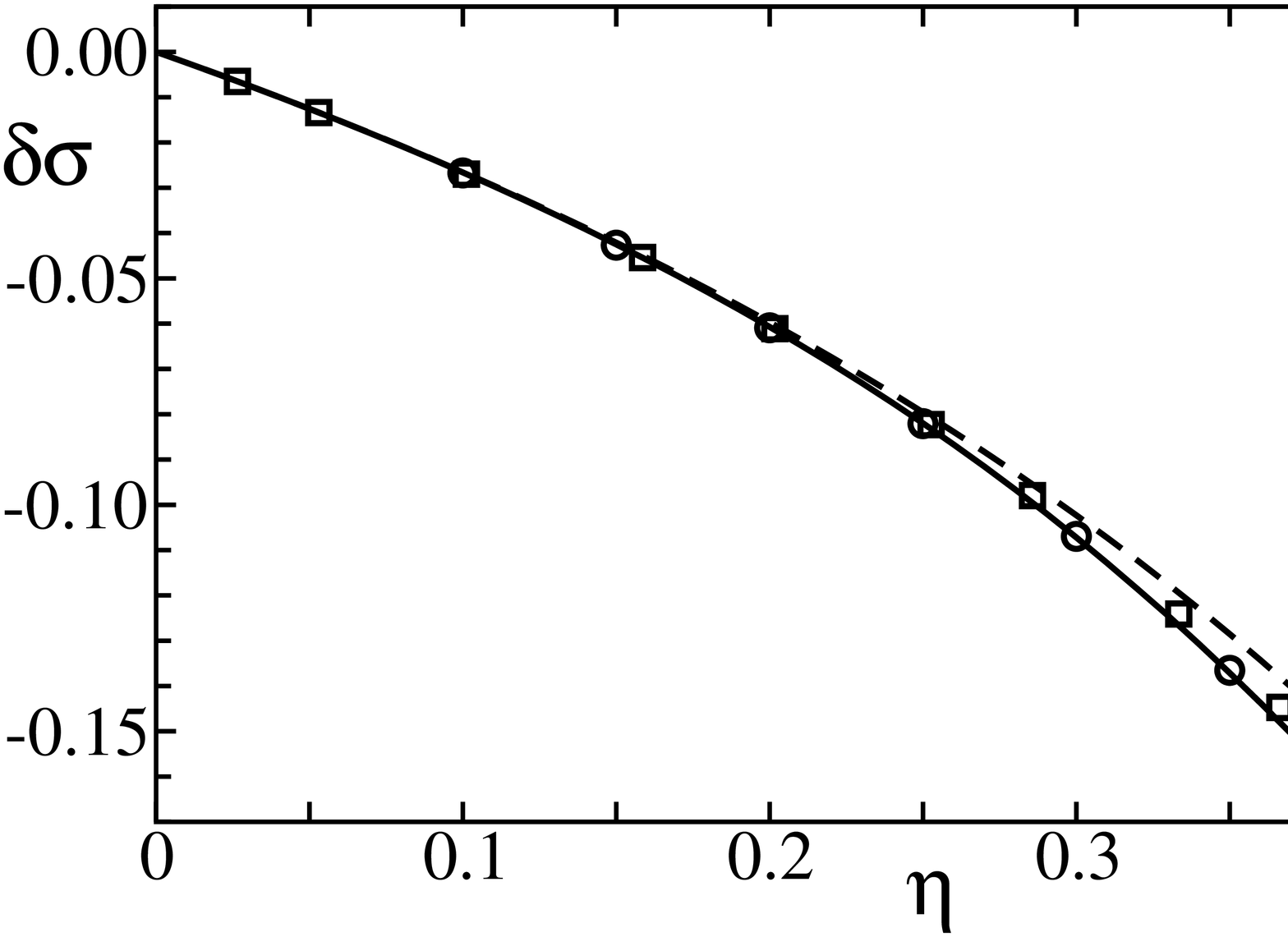}

\includegraphics[angle=270,width=250pt]{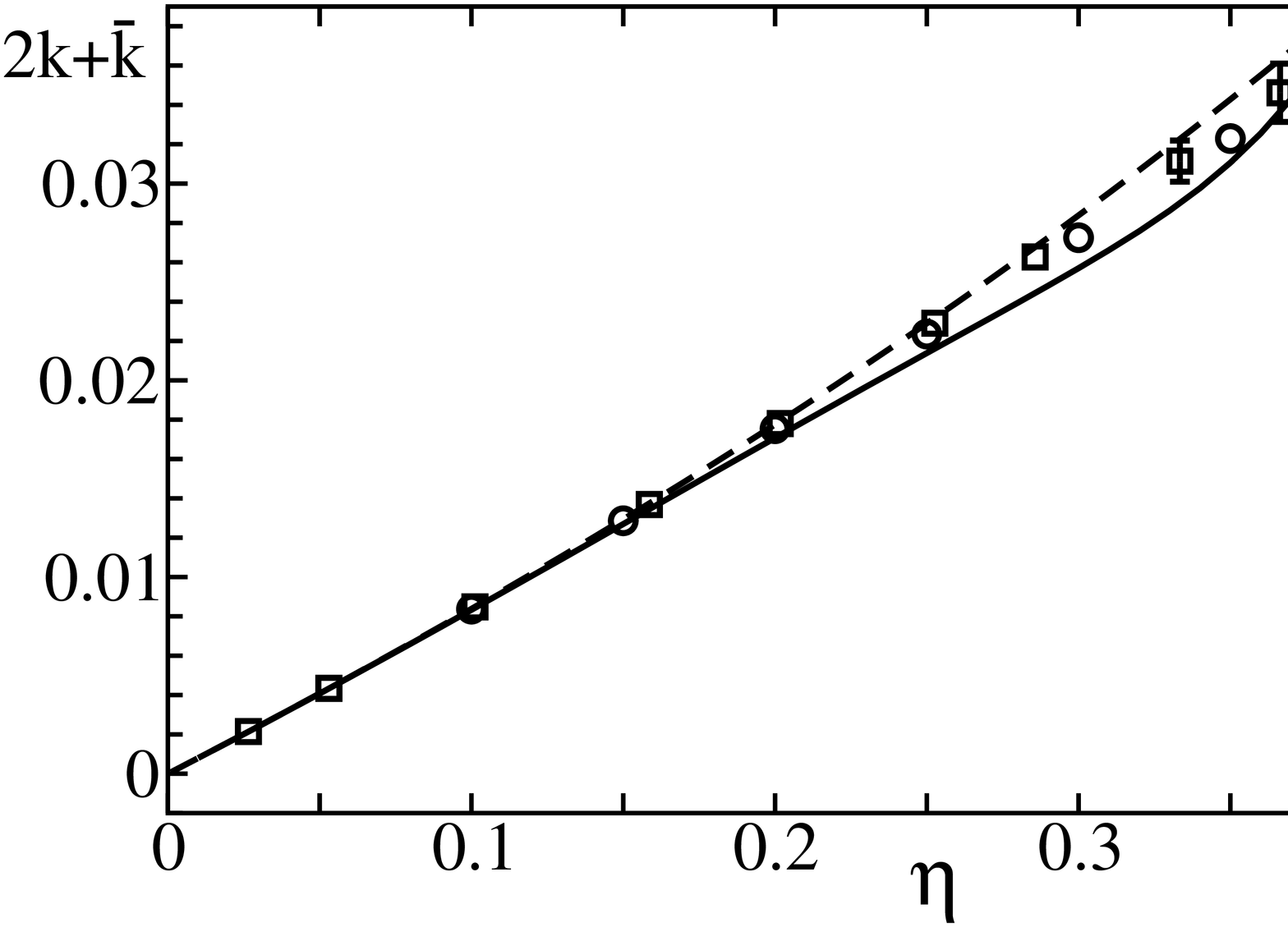}

\caption{Various curvature coefficients as a function of the fluid volume
fraction: (a) surface tension $\sigma$ (in units of $k_{\rm B} T / d^2$),
(b) Tolman length $\delta \sigma$ (in units of $k_{\rm B} T / d$)
and (c) the combination $2k + \bar{k}$ (in units of $k_{\rm B} T$).
The drawn lines are the results calculated using Eqs.(\ref{eq:sigma})-(\ref{eq:k_sph}),
transformed according to Eq.(\ref{eq:transformation}) so that the
radius is defined as that of the ``actual surface''. Circular symbols
are results from Bryk {\em et al.} \cite{Bryk03}; square symbols are the
computer simulation results by Laird {\em et al.} \cite{Laird10, Laird12};
the dashed line is the SPT result in Eq.(\ref{eq:SPT}).}
\label{fig:Fig1}
\end{figure}

The results for $\sigma$, $\delta \sigma$, and the combination $2k + \bar{k}$
are plotted in Figure \ref{fig:Fig1} as the solid lines. Also shown in Figure
\ref{fig:Fig1} are the calculations from Bryk {\em et al.} \cite{Bryk03}
(circular symbols), computer simulation results by Laird {\em et al.}
\cite{Laird10, Laird12} (square symbols) and Scaled Particle Theory (SPT)
\cite{SPT} (dashed lines), for which the expressions read:
\begin{eqnarray}
\label{eq:SPT}
\frac{\sigma \, d^2}{k_{\rm B} T}      &=& \frac{3 \, \eta (2 + \eta)}{2 \pi \, (1-\eta)^2} \,, \nonumber \\
\frac{\delta \sigma \, d}{k_{\rm B} T} &=& - \frac{3 \, \eta}{4 \pi \, (1-\eta)} \,,  \hspace*{27pt} {\rm (SPT)} \\
\frac{2k + \bar{k}}{k_{\rm B} T}       &=& - \frac{1}{4 \pi} \ln(1 - \eta) \,. \nonumber
\end{eqnarray} 
From the results in Figure \ref{fig:Fig1} it is concluded that the curvature
coefficients calculated using Eqs.(\ref{eq:sigma})-(\ref{eq:k_sph}) are consistent
with those obtained by Bryk {\em et al.} \cite{Bryk03}, although there seems to be
some small discrepancy for the combination $2 k + \bar{k}$ at larger volume fractions.
We come back to this point in the Discussion.

\begin{figure}
\centering
\includegraphics[angle=270,width=400pt]{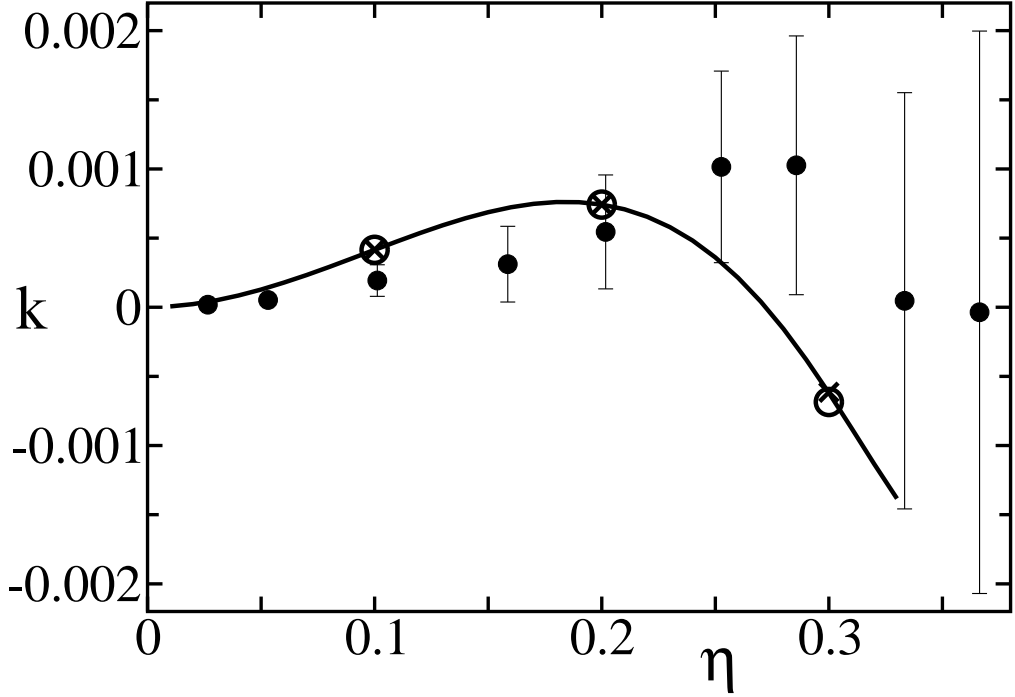}
\caption{Bending rigidity $k$ (in units of $k_{\rm B} T$)
as a function of the fluid volume fraction. The drawn line is the result
calculated from Eq.(\ref{eq:rigidity}); open circles and crosses are the
previous results listed in Table \ref{tb:Table2}; solid circles are the
computer simulation results by Laird {\em et al.} \cite{Laird12}.}
\label{fig:Fig2}
\end{figure}

\vskip 10pt
\noindent
{\bf Cylindrical interface}
\vskip 5pt
\noindent
The analysis for the cylindrical interface is more or less analogous to that
of the spherical interface, with one notable difference being that the weight
functions $w^c_{\alpha}(r_1,r_2)$ in Eq.(\ref{eq:w^c}) also need to be expanded 
in $1/R$:
\begin{eqnarray}
w^c_2(r_1,r_2)       &=& w^0_2(z_1-z_2) \, \left( 1 + \frac{d^2/4 - z_{12}^2}{16 \, R^2} + \ldots \right) \,, \\
w^c_{3}(r_1,r_2)     &=& w^0_3(z_1-z_2) \, \left( 1 + \frac{d^2/4 - z_{12}^2}{32 \, R^2} + \ldots \right) \,, \nonumber \\
w^c_{3'}(r_1,r_2)    &=& w^0_3(z_1-z_2) \, \left( 1 + \frac{3 (d^2/4 - z_{12}^2)}{32 \, R^2} + \ldots \right) \,, \nonumber \\
w^c_{\rm V}(r_1,r_2) &=& w^0_{\rm V}(z_1-z_2) \, \left( 1 + \frac{d^2/4 - z_{12}^2}{16 \, R^2} + \ldots \right) \,. \nonumber
\end{eqnarray}
Following the same procedure as for the spherical interface, the expressions for $\sigma$
and $\delta \sigma$ in Eqs.(\ref{eq:sigma}) and (\ref{eq:delta}) are recovered, and one
obtains as an expression for the bending rigidity $k$:
\begin{eqnarray}
\label{eq:rigidity}
\frac{k}{k_{\rm B} T} &=& - \frac{1}{4} \sum\limits_{\alpha} \int\limits_{-d/2}^{\infty}
\!\!\!\! dz_1 \int\limits_{0}^{\infty} \!\! dz_2 \; \frac{\partial \phi_0}{\partial n^0_{\alpha}(z_1)}
\, \rho_1(z_2) \, z_{12} \, w^0_{\alpha}(z_1-z_2) \\
&& + \frac{1}{4} \sum\limits_{\alpha, \beta} \int\limits_{-d/2}^{\infty}
\!\!\!\! dz_1 \int\limits_{0}^{\infty} \!\! dz_2 \; \frac{\partial^2 \phi_0}{\partial n^0_{\alpha}(z_1) \partial n^0_{\beta}(z_1)}
\, n^1_{\beta}(z_1) \, \rho_0(z_2) \, z_{12} \, w^0_{\alpha}(z_1-z_2) \nonumber \\
&& + \frac{1}{4} \sum\limits_{\beta} \int\limits_{-d/2}^{\infty}
\!\!\!\! dz_1 \int\limits_{0}^{\infty} \!\! dz_2 \; \frac{\partial^2 \phi_0}{\partial n^0_{\rm V}(z_1) \partial n^0_{\beta}(z_1)}
\, n^1_{\beta}(z_1) \, \rho_0(z_2) \, w^0_3(z_1-z_2) \nonumber \\
&& + \frac{1}{8} \sum\limits_{\alpha} \int\limits_{-d/2}^{\infty}
\!\!\!\! dz_1 \int\limits_{0}^{\infty} \!\! dz_2 \; \frac{\partial \phi_0}{\partial n^0_{\alpha}(z_1)}
\, \rho_0(z_2) \, (\frac{d^2}{4} - 3 z_{12}^2) \, w^0_{\alpha}(z_1-z_2) \nonumber \\
&& - \frac{1}{16} \int\limits_{-d/2}^{\infty} \!\!\!\! dz_1 \int\limits_{0}^{\infty} \!\! dz_2 \;
\frac{\partial \phi_0}{\partial n^0_3(z_1)} \, \rho_0(z_2) \, (\frac{d^2}{4} - z_{12}^2) \, w^0_3(z_1-z_2) \nonumber \\
&& + \frac{1}{4} \int\limits_{-d/2}^{\infty}
\!\!\!\! dz_1 \int\limits_{0}^{\infty} \!\! dz_2 \; \frac{\partial \phi_0}{\partial n^0_{\rm V}(z_1)}
\, \rho_1(z_2) \, w^0_3(z_1-z_2) \nonumber \\
&& + \frac{1}{2} \int\limits_{-d/2}^{\infty}
\!\!\!\! dz_1 \int\limits_{0}^{\infty} \!\! dz_2 \; \frac{\partial \phi_0}{\partial n^0_{\rm V}(z_1)}
\, \rho_0(z_2) \, z_{12} \, w^0_3(z_1-z_2) \,, \nonumber
\end{eqnarray} 
where $\rho_1(z)$ and $n^1_{\alpha}(z)$ are the same as in the spherical analysis.
It is noteworthy that since no reference to the location of the $z \!=\! 0$ plane
is made in this expression, the bending rigidity is {\em independent} of the
choice for the location of the radius $R$, i.e. $(k)_{\rm R-d/2} \!=\! k$. In this respect
the bending rigidity is a much more inherent property of the interface in question.
The result of the evaluation of the bending rigidity using Eq.(\ref{eq:rigidity})
is shown as the solid line in Figure \ref{fig:Fig2}. The open circles and crosses
are the previous results for $k$ listed in Table \ref{tb:Table2}. Also shown are
very recent computer simulation results by Laird {\em et al.} \cite{Laird12} (solid circles). 

Figure \ref{fig:Fig2} is the main result of this article. It shows that the
bending rigidity is definitively {\bf not equal to zero} in the context of FMT
theory for a hard sphere fluid near a hard wall and the Hadwiger Theorem
{\bf does not apply} in this case. We have shown this via three more or less
independent approaches which agree within numerical accuracy with each other.
A further corroboration of this result are the computer simulation results by 
Laird {\em et al.} \cite{Laird12}; although the agreement is not quantitative,
the shape of the volume fraction dependence of $k$ is strikingly similar.

Finally, we like to mention that by combining the expressions in Eqs.(\ref{eq:k_sph})
and (\ref{eq:rigidity}), an expression for the rigidity constant associated with
Gaussian curvature may be obtained:
\begin{eqnarray}
\frac{\bar{k}}{k_{\rm B} T} &=& \int\limits_{-d/2}^{\infty} \!\!\!\! dz \; 
z^2 \, \left[ \rho_0 \, \ln(\rho_0) - \rho_0 - \frac{\mu}{k_{\rm B} T} \, \rho_0 + \phi_0
+ \frac{p}{k_{\rm B} T} \, \Theta(z) \right] \\
&& + \sum\limits_{\alpha} \int\limits_{-d/2}^{\infty} \!\!\!\! dz_1 \int\limits_{0}^{\infty} \!\! dz_2 \;
z_1 \, \frac{\partial \phi_0}{\partial n^0_{\alpha}(z_1)}
\, \rho_0(z_2) \, z_{12} \, w^0_{\alpha}(z_1-z_2) \nonumber \\
&& + \int\limits_{-d/2}^{\infty} \!\!\!\! dz_1 \int\limits_{0}^{\infty} \!\! dz_2 \;
z_1 \, \frac{\partial \phi_0}{\partial n^0_{\rm V}(z_1)} \, \rho_0(z_2) \, w^0_3(z_1-z_2) \nonumber \\
&& - \frac{1}{4} \sum\limits_{\alpha} \int\limits_{-d/2}^{\infty} \!\!\!\! dz_1 \int\limits_{0}^{\infty} \!\! dz_2 \;
\frac{\partial \phi_0}{\partial n^0_{\alpha}(z_1)}
\, \rho_0(z_2) \, (\frac{d^2}{4} - 3 z_{12}^2) \, w^0_{\alpha}(z_1-z_2) \nonumber \\
&& + \frac{1}{8} \int\limits_{-d/2}^{\infty} \!\!\!\! dz_1 \int\limits_{0}^{\infty} \!\! dz_2 \;
\frac{\partial \phi_0}{\partial n^0_3(z_1)} \, \rho_0(z_2) \, (\frac{d^2}{4} - z_{12}^2) \, w^0_3(z_1-z_2) \,. \nonumber
\end{eqnarray} 
Note that $\bar{k}$ can be evaluated from the properties of the planar interface only;
a result that is consistent with similar expressions for the liquid-vapour interface \cite {Blokhuis93}.

\noindent
\section{Discussion}

\noindent
We have shown that the bending rigidity is not equal to zero in the context
of FMT theory for a hard sphere fluid near a hard wall and that the Hadwiger
Theorem does not apply in this case. Evidence for this conclusion is shown
in Figure \ref{fig:Fig2} where the results of three independent approaches
are shown to agree within numerical accuracy. Noteworthy is that the bending
rigidity changes sign from positive to negative as a function of increasing
fluid volume fraction. It is smaller than the rigidity constant associated
with Gaussian rigidity roughly by an order of magnitude.

The reduced magnitude of $k$ may certainly be partly responsible for the fact
that in a previous analysis \cite{Konig04} it was hard to distinguish it from
zero. Another possible source for the discrepancy may be due to a different
fit procedure used to extract the curvature coefficients from the radius
dependence of the surface tension and wall density. A comparison between
our analysis and the analysis in refs. \cite{Bryk03, Konig04, Konig05}
shows that while numerical results for $\sigma_s(R)$ agree to within a
high degree of accuracy \cite{Roth_communication}, the difference in
fit procedure leads to a fitted value for $2 k + \bar{k}$ that may differ
by as much as 10 \% (see Figure \ref{fig:Fig1}c). One may very well
speculate that the difference in fit procedure used may also have
consequences for the fitted value obtained for $k$.

The question now remains, what is the underlying physics of the Hadwiger
Theorem? The Hadwiger Theorem is not merely some abstract notion from
Mathematics and one should be able to understand more microscopically
when the conditions (i.e additivity) that lead to it are fulfilled.
To address this question, let us consider the general form of the mean-field
expressions for the surface tension in spherical (Eq.(\ref{eq:sigma_sphere}))
and cylindrical geometry (Eq.(\ref{eq:sigma_cylinder})) \cite{Blokhuis00}:
\begin{eqnarray}
\label{eq:sigma_general}
\sigma_s(R) &=& \int\limits \!\! dr \left( \frac{r}{R} \right)^{\!2} \, \Pi_s(r) \,, \\
\sigma_c(R) &=& \int\limits \!\! dr \left( \frac{r}{R} \right) \, \Pi_c(r) \,, \nonumber
\end{eqnarray} 
where $\Pi(r)$ depends on the distribution of the fluid density $\rho(r)$
in the interfacial region and may be referred to as the excess free energy
density or (the negative of) the excess lateral pressure \cite{Blokhuis00}.

\begin{figure}
\centering
\includegraphics[angle=270,width=400pt]{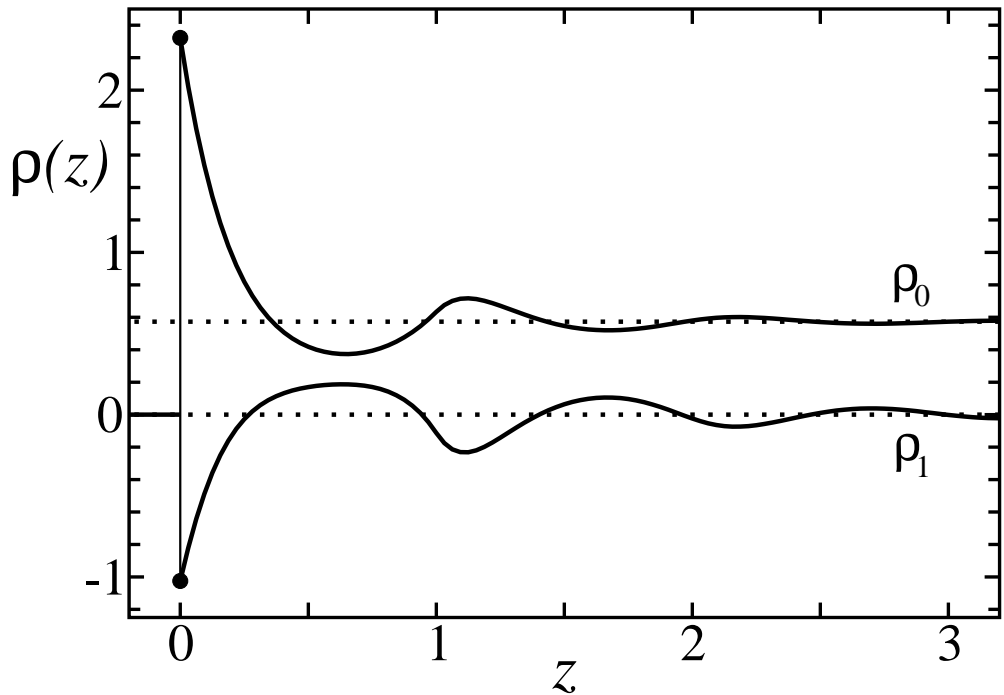}
\caption{Density profiles $\rho_0(z)$ (in units of $1/d^3$) and $\rho_1(z)$
(in units of $1/d^2$) as a function of $z$ (in units of $d$) for $\eta \!=\!$ 0.3.
The values at $z\!=\!0$ correspond to the pressure $\rho_0(0^+) \!=\! p / k_{\rm B} T$
and (twice) the surface tension $\rho_1(0^+) \!=\! 2 \sigma / k_{\rm B} T$;
cf. Eq.(\ref{eq:rho_s_w(R)}).}
\label{fig:Fig3}
\end{figure}

Now, if it is assumed that the lateral pressure is {\em independent}
of $R$, i.e. $\Pi(r) \!=\! \Pi_0(z)$, then the only radius dependence
in Eq.(\ref{eq:sigma_general}) is due to the geometric factors
$(r/R)^2$ and $(r/R)$. Therefore, we immediately conclude from
Eq.(\ref{eq:sigma_general}) that
\begin{eqnarray}
\sigma        &=& \int\limits \!\! dz \; \Pi_0(z) \,, \\
\delta \sigma &=& - \int\limits \!\! dz \; z \, \Pi_0(z) \,, \nonumber \\
\bar{k}       &=& \int\limits \!\! dz \; z^2 \, \Pi_0(z) \,, \nonumber
\end{eqnarray} 
and {\em the bending rigidity k is zero}. Furthermore, all the higher order
terms in the expansion in $1/R$ are absent. [It was already Helfrich himself who
derived these ``geometrical expressions'' in terms of progressing moments of the
excess lateral pressure \cite{HelfLH}.] These results correspond {\em precisely}
to the predictions of the Hadwiger Theorem so that we may conclude that the Hadwiger
Theorem corresponds to the statement:
\begin{eqnarray}
{\rm Hadwiger \,\,\, Theorem} \hspace*{27pt} \Longleftrightarrow \hspace*{27pt} \Pi(r) = \Pi_0(z) \,.
\end{eqnarray} 
This means that the Hadwiger Theorem applies when the fluid molecules do not
rearrange themselves when the curvature of the interface is changed. Certainly,
theoretical models may be constructed in which such a rearrangement does not occur,
but in general this is certainly not the case. To explore this curvature dependence,
we expand $\Pi_s(r)$ for a spherical interface in $1/R$:
\begin{equation}
\Pi_s(r) = \Pi_0(z) + \frac{\Pi_1(z)}{R} + \ldots \,.
\end{equation}
Szleifer and coworkers \cite{Szleifer} already showed that the
bending rigidity $k$ is then expressed as
\begin{equation}
k = \frac{1}{2} \int\limits \!\! dz \; z \, \Pi_1(z) \,,
\end{equation}
which explicitly demonstrates the conclusion that $k$ results from the
(possible) rearrangement of molecules when the curvature of the interface
is changed. An example of such a rearrangement of molecules as described
by the density profile $\rho_1(z)$ is shown in Figure \ref{fig:Fig3} for
$\eta \!=\!$ 0.3.

Now, one could argue that the vanishing of the bending rigidity is simply
a matter of length-scale \cite{Konig04}. The length-scale associated with
the molecular rearrangement due to curvature is the width of the interfacial
region $\xi$, which is small compared any to {\em macroscopic} length-scale
unless the system is critical \cite{RW} or when a macroscopic wetting
layer is present \cite{Evans03, Evans04, Evans05}. However, the same
argument would apply to {\em all} the curvature coefficients and in particular
to the rigidity constant associated with Gaussian curvature which scales
similarly to the bending rigidity.
\vskip 15pt
\noindent
{\Large\bf Acknowledgment}
\vskip 5pt
\noindent
This article was inspired by a wonderful presentation by Roland Roth at the
Jim Henderson retirement symposium. Further discussions with him and Bob Evans
were greatly appreciated. I would also like to thank Brian Laird for communicating
his simulation results prior to publication.


\begin{thebibliography}{}
\bibitem{Note}
The absence of this term leads to the striking result in Figure 3 in ref. \cite{Konig04}.
It is, however, unrelated to the question whether the bending rigidity vanishes or not.
\end{thebibliography}
\end{document}